\documentclass[12pt]{iopart}
\usepackage{graphics}
\usepackage{graphicx}
\usepackage{bm}
\usepackage{amssymb}
\usepackage{amsfonts}
\usepackage{amsthm}
\usepackage{bm}
\begin{document}
\title{Entanglement and coherence  in a spin-$s$  $XXZ$ system under non-uniform fields}

\author{E. R\'{\i}os$^{1}$, R.\ Rossignoli $^{2}$, N.\ Canosa$^{2}$}

\address{$^{1}$Departamento de Ingenier\'ia Qu\'imica,
    Universidad Tecnol\'ogica Nacional, Facultad Regional Avellaneda, C.C. 1874, Argentina\\
    $^{2}$ IFLP-Departamento de F\'{\i}sica-FCE, Universidad Nacional de La Plata,
    C.C.\ 67, 1900 La Plata, Argentina}

\begin{abstract}
We investigate entanglement and coherence in an $XXZ$ spin-$s$ pair immersed in
a non-uniform transverse magnetic field. The  ground state and thermal 
entanglement phase diagrams are analyzed in detail  in both 
the ferromagnetic and antiferromagnetic cases. It is  shown that a
non-uniform field enables to control the energy levels and the entanglement of
the corresponding eigenstates, making it possible  to entangle the system for any 
value of the exchange couplings, both at zero and finite temperatures. Moreover, the limit
temperature for entanglement is shown to depend only on the difference
$|h_1-h_2|$ between the fields applied at each spin, leading for $T>0$ to a
separability stripe in the $(h_1,h_2)$ field plane such that the system becomes
entangled above a threshold value of $|h_1-h_2|$. These results are demonstrated to be
rigorously valid for any spin $s$. On the other hand, the relative entropy of
coherence in the standard basis, which coincides with the ground state
entanglement entropy at $T=0$ for any $s$, becomes non-zero for any value of
the fields at $T>0$, decreasing uniformly for sufficiently high $T$. A special
critical point arising at $T=0$  for non-uniform fields in the ferromagnetic case is also
discussed.

\end{abstract}
\maketitle
\section{Introduction}

The theory of quantum entanglement has provided a useful and novel  perspective
for the analysis of correlations and quantum phase transitions  in interacting
many body systems \cite{Am.08,ECP.10,ON.02,VLRK.03,VMDC.04}. At the same time, it is
essential for determining the capability  of such systems for performing
different quantum information tasks \cite{NC.00,HR.06,DVz}. More
recently,  a general theory of quantum resources, similar to that of
entanglement but based on the degree of coherence of a quantum system with respect
to a given reference basis, was proposed \cite{V.14,B.14,G.14,JMM.15}. 
Thus, entanglement and coherence  provide a means  to capture the degree of quantumness  of a given
quantum system. 

In particular, spin systems constitute  paradigmatic examples of strongly interacting many
body systems  which enable to study in detail the previous issues, providing
at the same time a convenient scalable scenario for the implementation of
quantum information protocols. Interest on spin systems has been recently
enhanced by the significant advances in  control techniques of quantum systems,
which have permitted the simulation of interacting spin models with different
type of couplings by means of  trapped ions, Josephson junctions or cold atoms
in optical lattices \cite{PC.04,GA.14,BR.12,SR.15,B.16,LS.12}.

Accordingly, interacting spin systems have been the object of several 
relevant studies. Entanglement and discord-type correlations \cite{OZ.01,Mo.12,A.16,RCC.10}  
 in spin pairs and chains with Heisenberg couplings under uniform fields were intensively investigated, 
 specially for spin $1/2$ systems 
\cite{ON.02,Ar.01,G.01,W.02,WFS.01,KS.02,GBF.03,CR.04,PVMDC.05,RC.05,D.08,CRC.10,SK.13}.
The effects of  non-uniform 
fields have also received attention, mostly for spin $1/2$ systems 
\cite{SCC.03, AK.05,ZL.05,HYP.07,HL.10,Guo.11,AA.06,Zh.11}, although some results 
for  higher spins in  non-uniform fields are also available \cite{EA.10,GG.10,Hu.14}.

The aim of this work is to analyze in detail the effects of a non-uniform magnetic field
on the entanglement and coherence  of a spin-$s$ pair interacting through an
$XXZ$ coupling, both at zero and finite temperature. We  examine the interplay
between the non-uniform magnetic field and temperature and their role to
control quantum correlations. We  also study the critical behaviour and the
development of different phases as the spin increases,  when the field,
temperature, and coupling anisotropy are varied. Analytical rigorous results
are also  provided. In particular,  the $T=0$ phase diagram will be 
characterized by ground states of definite magnetization $M$, all reachable
through non-uniform fields for any value of the  couplings,  with
entanglement decreasing with increasing $|M|$. Special critical points will be
discussed. On the other hand,  the limit temperature for entanglement will be
shown to depend, for any value of $s$, only on the difference between the fields 
applied at each spin, leading to a thermal phase diagram characterized by a
separability stripe in field space. Finally, we will analyze the relative
entropy of coherence \cite{B.14} in the standard basis, 
which coincides here exactly with the
entanglement entropy at $T=0$ but departs from entanglement as $T$ increases.

The model and results are presented in sections II--IV, starting with the basic
spin $1/2$ case and considering then the $s=1$  and the general spin-$s$ cases.
Conclusions are finally given in section V.

\section{Model and the spin 1/2 case}
We consider a spin $s$ pair interacting through  an $XXZ$-type  coupling,
immersed in a transverse magnetic field $\bm{h}$
not necessarily uniform. The Hamiltonian can be written as
 \begin{equation}
H=-h_{1}s_{1}^{z}-h_2 s_2^z+J(s_{1}^{x} s_{2}^x+s_1^y s_{2}^y)+J_z s_{1}^{z} s_{2}^z\,,
\label{H1}
\end{equation}
where $s^{\mu}_{i}$ ($\mu=x,y,z$) denote the (dimensionless) spin operators at
site $i$ and $J$, $J_z$ the exchange couplings, with $J_z/J$ the anisotropy
ratio. This Hamiltonian commutes with the total spin  along the $z$ axis,
$S_z=s^z_1+s^z_2$, having then eigenstates with definite magnetization $M$
along $z$. Without loss of generality, we will set in what follows $J>0$,  as
its sign can be changed by a local rotation of angle $\pi$ around the $z$ axis
of one of the spins, which will not affect the energy spectrum nor the
entanglement of its eigenstates. The ferromagnetic (FM) case $J<0$, $J_z<0$ is
then equivalent to  $J>0$, $J_z<0$.

We also remark that a  Hamiltonian with an additional Dzyaloshinskii-Moriya
coupling along $z$ \cite{DM}, $H'=H+D\sum_{i} (s_{i}^{x} s_{i+1}^y- s_i^y s_{i+1}^x)$,
can be transformed back exactly into a Hamiltonian (\ref{H1}) with
$J\rightarrow J'=\sqrt{J^2+D^2}$, by means of a rotation of angle $\phi=\tan^{-1}(D/J)$   
around the $z$ axis at the second spin \cite{DM2}.  Hence, its
spectrum and entanglement properties will also coincide {\it exactly} with
those of Eq.\ (\ref{H1}) for $J\rightarrow J'$.

We first review the $s=1/2$ case, providing a complete study with analytical 
results and including coherence in the standard basis,  
which allows to understand more easily the general spin $s$ case, considered in the
next subsections. Entanglement and discord-type  correlations under non homogeneous fields in a 
spin $1/2$ pair were  studied in \cite{SCC.03,HL.10} for an $XX$-type coupling, in \cite{AK.05} 
for an isotropic coupling, in \cite{ZL.05,Guo.11} for an $XXZ$ coupling  
and in \cite{AA.06} for an $XYZ$ coupling. 

\subsection{The spin $1/2$ pair}

Using qubit notation, the eigenstates of the Hamiltonian (\ref{H1}) for $s=1/2$
are the separable aligned states  $|00\rangle\equiv |\uparrow\uparrow\rangle$
and  $|11\rangle\equiv|\downarrow\downarrow\rangle$,  with magnetization
$M=\pm 1$ and energies
\begin{equation}E_{\pm 1}= \mp\frac{1}{2}(h_1+h_2)+\frac{1}{4}J_z\,,\label{E1}\end{equation}
and the entangled $M=0$  states $\left|\Psi^{\pm}\right\rangle
=\cos\alpha_{\pm} |01\rangle+\sin\alpha_{\pm}|10\rangle$,   with  energies
\begin{equation}
E_{0}^{\pm}=\pm\frac{1}{2}\Delta-\frac{1}{4}J_z\,,\;\;
 \Delta=\sqrt{(h_{1}-h_{2})^{2}+J^{2}}\,,\label{E0}\end{equation}
and $\tan\alpha_{\pm}=\frac{h_1-h_2\pm \Delta}{J}$. The  concurrence \cite{WW}
of these states   is given by
\begin{equation}
C\left(\left|\Psi^{\pm}\right\rangle\right)=|\sin 2\alpha_\pm|=J/\Delta\,,
\label{con1}
\end{equation}
and is a decreasing function of $|h_1-h_2|/J$. Their entanglement entropy,
$S=-{\rm Tr}\rho_i\log_2\rho_i$ with $\rho_i$ the reduced state of one of the
spins, can then be obtained as
\begin{equation}
{\textstyle S=-\sum_{\nu=\pm}p_\nu\log_2
 p_\nu\,,\;\;p_{\pm}=\frac{1\pm\sqrt{1-C^2}}{2}}\,, \label{S}\end{equation}
and is an increasing function of $C$.  In the uniform case $h_{1}=h_2$,
$\left|\Psi^{\pm}\right\rangle$ become the Bell states
$\frac{\left|01\right\rangle \pm \left|10\right\rangle}{\sqrt{2}}$ and
$S(|\Psi^{\pm})=C(|\Psi^{\pm}\rangle)=1$.

Through a non-uniform field it is then possible to tune the entanglement of the
$M=0$ eigenstates,  decreasing it by applying a field difference. On the other
hand, such difference also decreases the energy $E_0^-$ of $|\Psi^{-}\rangle$
and increases that of $|\Psi^+\rangle$, without affecting that of the aligned
eigenstates if the average field is kept constant,  enabling to have the
entangled state $|\Psi^-\rangle$ as a non-degenerate ground state (GS) for any
value of $J$ or $J_z$. A similar effect can be obtained by increasing $J_z$,
which increases the gap between the entangled and the aligned states, in this
case without affecting their concurrence.

\begin{figure}[htb]
 \centerline{\includegraphics[scale=0.60]{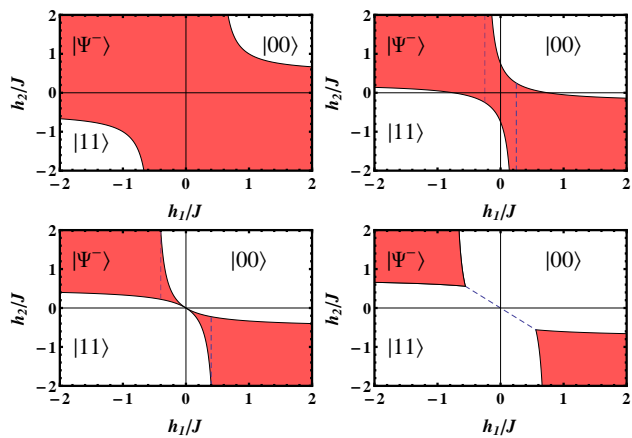}}
\caption{Ground state phase diagram for the spin $1/2$ pair. Top:
antiferromagnetic (AFM) case  $J_{z}=J$ (left) and ferromagnetic (FM)-type case
$J_{z}=-J/2$ (right). Bottom: FM cases $J_{z}=-J$ (left) and
$J_{z}=-\frac{3}{2}J$ (right).} \label{f1}
\end{figure}

Eqs.\ (\ref{E1})--(\ref{E0}) then lead to the phase diagrams of Fig.\ \ref{f1}.
For  clarity we have considered the whole field plane, although the diagrams
are obviously symmetric under reflection from the $h_1=h_2$ line (and spectrum
and entanglement also from the $h_1=-h_2$ line). The GS will be either the
entangled state $|\Psi^-\rangle$ (red sector) or one of the aligned  states
($|00\rangle$ if $h_1+h_2>0$ or $|11\rangle$  if $h_1+h_2<0$, white sectors),
with  $|\Psi^-\rangle$ a non-degenerate GS ($E_0^-<E_{\pm 1}$) if and only if
\begin{equation}|h_1+h_2|<J_z+\sqrt{J^2+(h_1-h_2)^2}.\label{c0}\end{equation}
This equation is equivalent to the following conditions:
\begin{eqnarray}
\!\!\!\!(h_1-{\textstyle\frac{J_z}{2}})(h_2-{\textstyle\frac{J_z}{2}})&<&
{\textstyle\frac{J^2}{4}}\,,\;\;(h_1+h_2\geq 0,\;h_1>{\textstyle\frac{J_z}{2}})\label{c1}\\
\!\!\!\!(h_1+{\textstyle\frac{J_z}{2}})(h_2+{\textstyle\frac{J_z}{2}})&<&{\textstyle\frac{J^2}{4}}\,,
\;\;(h_1+h_2\leq 0,\;h_1<\!-{\textstyle\frac{J_z}{2}})\,.
\label{c2}
\end{eqnarray}
which show that the borders of the entangled sector are displaced  hyperbola
branches.

In the AFM case $J_z>0$, the diagram has the form of the top left panel. Here
the GS is entangled at zero field and  if one of the fields is sufficiently
weak ($|h_1|<J_z/2$) the GS remains entangled for any value of the other field.
However, in the FM case $J_z<0$ two distinct diagrams can arise (top and bottom
right panels), separated by the limit diagram of the bottom left panel
($J_z=-J$). If $-J<J_z<0$ (top right), the system is still entangled at zero
field but now if one of the fields is sufficiently weak ($|h_1|<|J_z|/2$, dashed
vertical lines) entanglement is  confined to a finite interval of the other
field.  Control of just one field then allows to switch entanglement on and off
for any value of the other field.

On the other hand, if $J_z<-J$,  the GS is aligned  for {\it any} uniform
non-zero field, and  $|\Psi^-\rangle$ becomes GS only above a {\it threshold}
value of the field difference, $|h_1-h_2|>\sqrt{J_z^{2}-J^{2}}$ (Eq.\
(\ref{c0})), within the limits determined by Eqs.\ (\ref{c0})--(\ref{c2}).
These limits imply that the sign of the field at each site must be {\it
different},  as seen in the bottom right panel.  Hence, GS entanglement is in
this case switched on (rather than destroyed) by field application, provided it
has  opposite signs at each spin. In addition, a GS transition between the
aligned states $|11\rangle$ and $|00\rangle$ takes place at the line $h_1=-h_2$
for  $|h_1-h_2|<\sqrt{J_z^2-J^2}$, with the GS degenerate in this interval
along this line.

\begin{figure}[htb]
 \centerline{\includegraphics[scale=0.8]{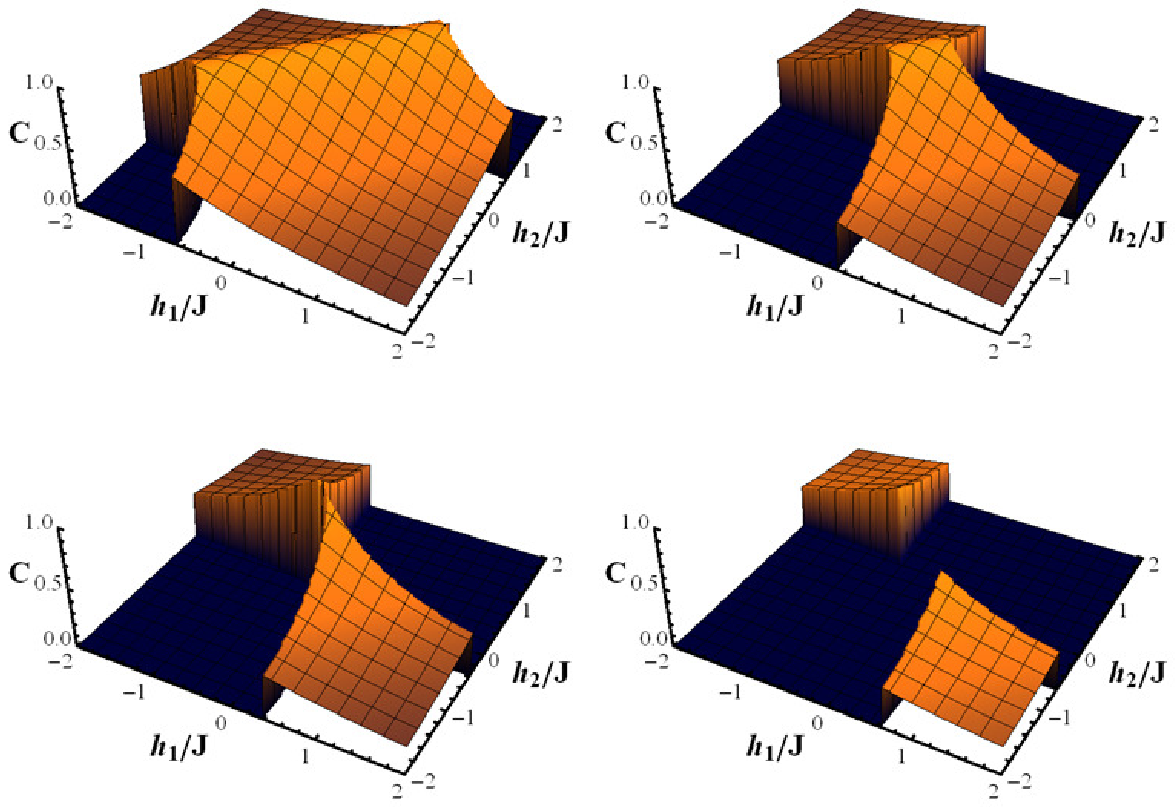}}
\caption{Concurrence of the GS as a function of the magnetic fields for the
cases of Fig.\ \ref{f1}. Top:  $J_{z}=J$ (left) and  $-J/2$
(right). Bottom: $J_{z}=-J$ (left) and $-\frac{3}{2}J$ (right). } \label{f2}
\end{figure}

The GS concurrence  for the same cases of Fig.\ \ref{f1} is depicted on Fig.\
\ref{f2}. As seen in the top  panels,  the maximum $C=1$ is reached for
$h_1=h_2=h$ provided   $J_z>-J$ and $|h|<\frac{J+J_z}{2}$. For $J_z<-J$, the
maximum value is  $C=J/|J_z|<1$, attained  at the edges
$h_1=-h_2=\pm\frac{1}{2}\sqrt{J_z^{2}-J^{2}}$ of the entangled sector.

\vspace*{-0.5cm}

\subsubsection{Thermal entanglement}
Let us now consider a finite temperature $T$. As $T$ increases from 0, an
entangled GS will become mixed with other excited states, leading  to a
decrease of the entanglement which will  vanish beyond a limit temperature.
However, if the GS is separable, the thermal state  can become entangled for
$T>0$  (below some limit temperature) due to the presence of entangled  excited
states, implying that the entanglement phase diagram for $T>0$ may {\it differ}
from that at $T=0$ even for low $T$.

In the present case the thermal state $\rho_{12}=Z^{-1}e^{-\beta H}$, with
$Z={\rm Tr}\,e^{-\beta H}$ the partition function and $\beta=1/kT$,   has in
the standard basis  the form
\begin{equation}
\rho_{12}=\left(\begin{array}{cccc}p_{+}&0&0&0
\\0&q_+&w&0\\0&w&q_-&0\\0&0&0&p_{-}\end{array}\right)\,,
 \label{r12}\end{equation}
where $p_{\pm}=Z^{-1}e^{-\beta E_{\pm 1}}$, $q_{\pm}=Z^{-1}e^{\beta
J_z/4}(\cosh\frac{\beta \Delta}{2}\pm\frac{h_{1}-h_{2}}{\Delta}\sinh\frac{\beta
\Delta}{2})$ and $w=-Z^{-1}e^{\beta J_z/4}\frac{J}{\Delta}\sinh\frac{\beta
\Delta}{2}$. Its concurrence \cite{WW} is then given by
\begin{eqnarray}
C&=&2{\rm Max}[|w|-\sqrt{p_{+}p_{-}}, 0]={\textstyle 2Z^{-1}{\rm
Max}[\frac{J}{\Delta}e^{\beta J_z/4}\sinh\frac{\beta \Delta}{2}-e^{-\beta
J_z/4},0]}\,.
\end{eqnarray}
Thus, for $T>0$ $\rho_{12}$ is entangled if and only if 
\begin{equation}
{\frac{J}{\Delta} e^{\beta J_z/2}\sinh\frac{\beta \Delta}{2}>1}\,.\label{JzC}
\end{equation}

Eq.\ (\ref{JzC}) implies a limit temperature for entanglement that will  depend
on $J_z$, $J$ and $|h_1-h_2|$ only. It also implies  a {\it threshold value} of
$J_z$ for entanglement at {\it any} $T>0$,
\begin{equation}
J_z>2 kT\ln \frac{\Delta/J}{\sinh\frac{\beta
\Delta}{2}}=-\Delta+4kT\ln\frac{\Delta/J}{1-e^{-\beta \Delta}}\,.\label{jzc}
\end{equation}
Hence,  it is always possible to entangle the thermal state by increasing
$J_z$,  since  it will effectively cool down the system to the state
$|\Psi^-\rangle$, as previously stated.

The same effect occurs if the field difference $|h_1-h_2|$ is increased. The
left hand side of Eq.\ (\ref{JzC}) is an {\it increasing} function of $\Delta$ and
hence of  $|h_1-h_2|$ for any $T>0$ and $J_z$, so that at any $T>0$ there will
also exist a {\it  threshold value} $h_c$  of the field difference $|h_1-h_2|$
above which the thermal state  will become entangled:
\begin{equation} |h_1-h_2|> h_c(T,J,J_z)\label{hcz}\,.\end{equation}

\begin{figure}[htb]

 \centerline{\includegraphics[scale=0.5]{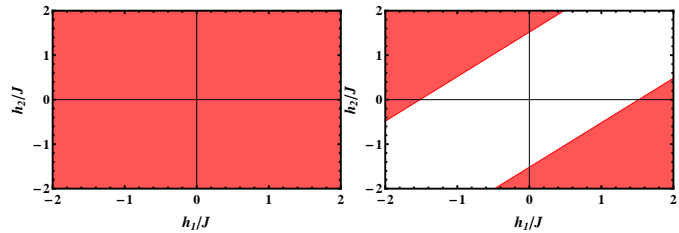}}
\caption{Thermal entanglement phase diagram for the spin $1/2$ pair at
$kT=J/2$.  Left: AFM case  $J_{z}=J$.  Right: FM case $J_{z}=-J/2$. Red sectors
indicate entanglement. The whole plane remains entangled for $0<kT<0.91 J$ if
$J_z=J$ and $0<kT< 0.335 J$ if $J_z=-J/2$.   Above these temperatures, a
separability stripe $|h_1-h_2|\leq h_c(T,J,J_z)$ arises. For $J_z< -J$ the
separability stripe arises  for any $T>0$. } \label{f3}
\end{figure}

Eq.\ (\ref{hcz}) gives rise to a {\it separability stripe} $|h_1-h_2|\leq
h_c(T,J,J_z)$, as depicted in Fig.\ \ref{f3}. Here
$h_c(T,J,J_z)=\sqrt{\Delta_c^2-J^2}$, with $\Delta_c=2kTf^{-1}(\frac{2kT}{J}
e^{-\beta J_z/2})$ and $f^{-1}$ the inverse of the increasing function
$f(x)=\sinh x/x$ ($x>0$).

Eq.\ (\ref{hcz})  implies that the thermal entanglement phase diagram in the
field plane  {\it differs}  from the $T = 0$ phase diagram even for  small
temperatures $T>0$, as it is determined just by the field difference
$|h_1-h_2|$. Entanglement will be  turned on in $T=0$ separable sectors outside
the stripe as soon as $T$ becomes finite. In particular, for $J_z>-J$, Eq.\
(\ref{jzc}) shows  that in contrast with the $T=0$ case, the {\it whole} $h_1,
h_2$ plane becomes entangled for $0 < T < T_c$, with $T_c(J,J_z)$ determined by

\begin{equation} e^{\beta_c J_z/2}\sinh{\textstyle\frac{\beta_c J}{2}}=1\,,
\label{Tcjz}\end{equation}
such that  $h_c(T,J,J_z)=0$ if $T<T_c(J,J_z)$. The separability stripe arises
then for $T>T_c(J,J_z)$.  For $J_z\rightarrow-J$, $T_c\rightarrow 0$ while if
$J_z=0$, $kT_c=\frac{1}{2}J/{\rm arcsinh}\,1\approx 0.567 J$.

However,  for $J_z<-J$  a separability stripe  will be present for all $T>0$,
with  $h_c(T,J,J_z)\rightarrow \sqrt{J_z^2-J^2}$ for $T\rightarrow 0^+$. The
thermal phase diagram in the field plane is then  characterized, for any value
of $J_z$,   by a separability stripe whose width increases with increasing $T$,
and vanishes for $J_z>-J$ if  $T<T_c(J,J_z)$.

\begin{figure}[htb]
    \centerline{\includegraphics[scale=0.8]{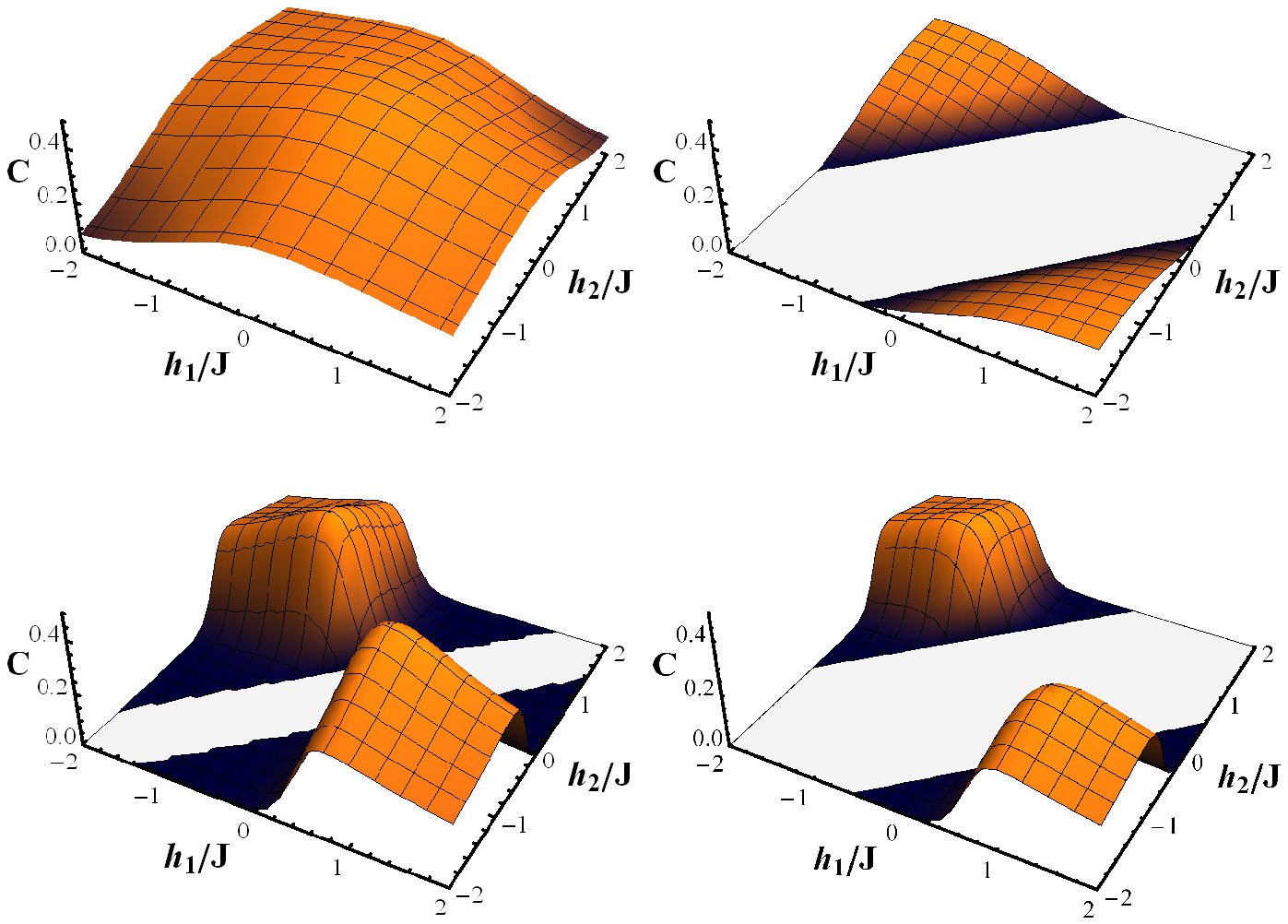}}
\caption{ Concurrence as a function of the magnetic fields $h_1$ and $h_2$ at
finite temperature. Top panels: AFM case  $J_{z}=J$ (left) and FM case
$J_{z}=-J/2$ (right) at temperature $kT=J/2$. Bottom panels: Same diagrams for the FM
cases  $J_{z}=-J$ (left) and $J_{z}=-\frac{3}{2}J$ (right) at  $kT=0.05 J$. }
\label{f4}
\end{figure}

\begin{figure}[htb]
    \centerline{\includegraphics[scale=.8]{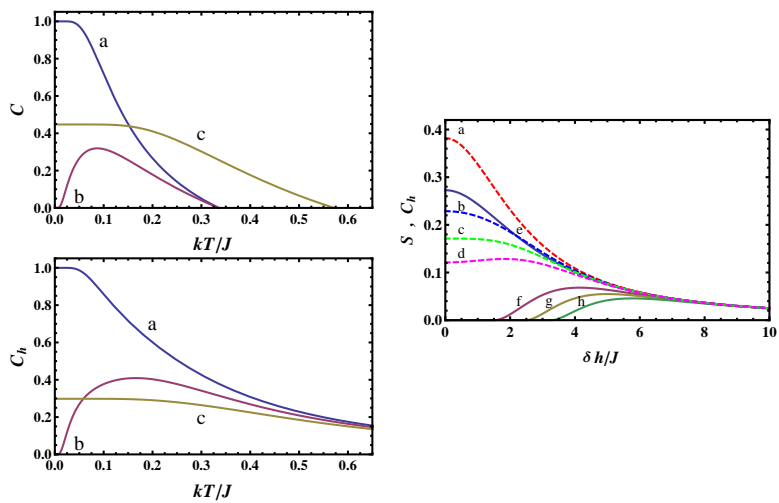}}
\caption{Left: Concurrence (upper panel) and relative entropy of coherence in
the standard basis (lower panel) as a function of temperature for $J_z=-J/2$ at
$h_1=h_2=0$ (a), $h_1=h_2=0.3 J$ (b) and $h_1=-h_2=J$ (c). The reentry of
entanglement for $T>0$ in case (b) is clearly seen, with the limit temperature
for entanglement independent of the field if $h_1=h_2$ (cases a,b) and
increasing with increasing values of $|h_1-h_2|$ (case c),  as follows from
Eq.\ (\ref{JzC}). In contrast, the coherence remains non-zero $\forall$ $T$,
decreasing uniformly as $(J/kT)^2$ for high $T$ (Eq.\ \ref{cha}) and
approaching the entanglement entropy (\ref{S})  for  $T\rightarrow 0$. Right:
Entanglement of formation $S$ (solid lines) and relative entropy of coherence
(dashed lines) as a function of the field difference $\delta h=|h_1-h_2|$ at
fixed temperature $kT=J/2$ for $J_z/J=1$ (a,e), $-1/2$ (b,f), $-1$ (c,g) and
$-3/2$ (d,h), at $h_1=-h_2$. As $|\delta h|$ increases,  all curves coalesce
and become independent of $J_z$, approaching the entanglement entropy of the GS
$|\Psi^-\rangle$ determined by Eq.\ (\ref{con1}).
        } \label{f5}
\end{figure}

The thermal concurrence is shown in Fig.\ \ref{f4}. It is verified that  it is
strictly zero just within the separability stripe (\ref{hcz}), becoming
small but non-zero in the $T=0$ separable regions outside it (dark blue in Fig.\ \ref{f4}). 
Nonetheless, such reentry of entanglement for $T>0$ can become quite noticeable in some
cases, as seen in the top panel of Fig.\ \ref{f5}. It is also verified that
through non-uniform fields it  becomes possible to preserve  entanglement up
to temperatures higher than those in the uniform case (which lies at
the center of the separability stripe), as also seen in Fig.\ \ref{f5}.

 \begin{figure}[htb]
    \centerline{\includegraphics[scale=0.8]{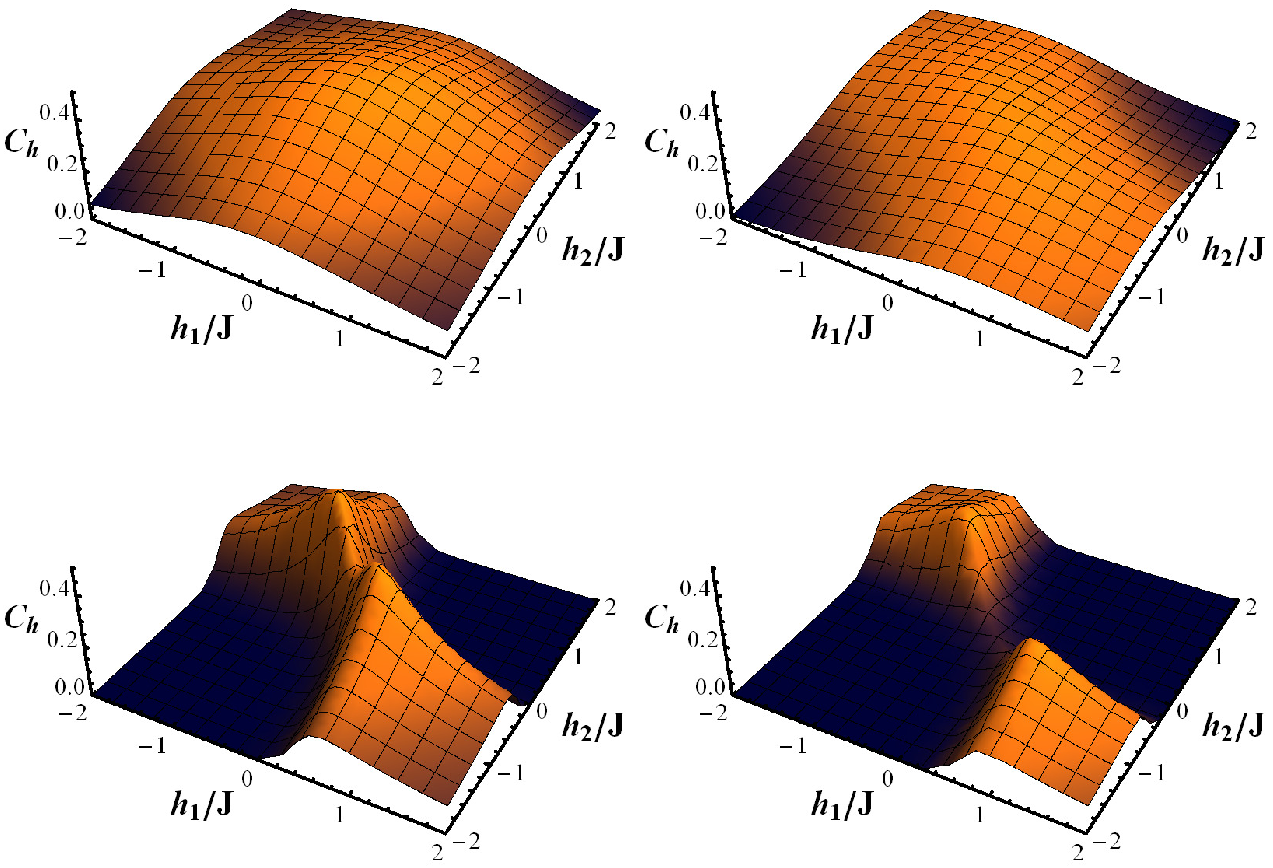}}
\caption{Relative entropy of coherence in the standard basis as a function of
the transverse non-uniform fields $h_1$ and $h_2$ at finite temperatures, for
the same cases of Fig.\ \ref{f4}.  Top panels: AFM case  $J_{z}=J$ (left) and
FM case $J_{z}=-J/2$ (right) for $kT=J/2$. Bottom panels: FM cases  $J_{z}=-J$ (left)
and $J_{z}=-\frac{3}{2}J$ (right) for $kT=0.1 J$.} \label{f6}
 \end{figure}

\subsubsection{Coherence}
We now analyze the coherence  of the thermal state (\ref{r12})
with respect to the standard product basis $\{|00\rangle, |01\rangle,
|10\rangle,|11\rangle\}$. This quantity can be measured through the relative
entropy of coherence  \cite{B.14}, defined as
\begin{equation}
C_h(\rho_{12})=S(\rho_{12}||\rho_{12}^{\rm diag})=S(\rho_{12}^{\rm diag})-S(\rho_{12}),\label{Ch}
\end{equation}
where $S(\rho)=-{\rm Tr} \rho \log_2 \rho$ is the von Neumann entropy and
$\rho_{12}^{\rm diag}$ its diagonal part in the previous basis. It is a measure
of the strength of the  off-diagonal elements in this basis, and would
obviously vanish if $J=0$. It will be here driven just by the coefficient $w$
in (\ref{r12}). A series expansion of (\ref{Ch})  for $|w|\ll q_{\pm}$ in
(\ref{r12}) leads in fact  to $C_h(\rho_{12}) \approx
\frac{\log_2(q_+/q_-)}{q_+-q_-}\,w^2$. The exact expression  is
$C_h(\rho_{12})=-\sum_{\nu=\pm}(q_\nu\log_2 q_\nu-p^0_\nu\log_2 p^0_\nu)$,
where $p^0_\pm=Z^{-1}e^{-\beta E_0^\pm}$.

In the zero temperature limit, $S(\rho_{12})$ vanishes while $S(\rho_{12}^{\rm
diag})$ and hence $C_h(\rho_{12})$ become the {\it entanglement entropy} $S$ of
the GS, Eq.\ (\ref{S}), since  the standard basis is here the Schmidt basis for
$|\Psi^-\rangle$. However, for $T>0$, $C_h(\rho_{12})$ becomes everywhere
non-zero due to the non-vanishing weight of the entangled states
$|\Psi^\pm\rangle$, as seen in Figs.\ \ref{f5} and \ref{f6}. In fact, for
$kT\gg {\rm max}[J,J_z,h_1,h_2]$, a series expansion leads to  the  asymptotic
expression
\begin{eqnarray}
C_h(\rho_{12})&\approx &{\textstyle\frac{1}{16\ln
2}\left(\frac{J}{kT}\right)^2\left[1+\frac{J_z}{4kT}
-\frac{3[(h_1+h_2)^2+J^2]+(h_1-h_2)^2}{48(kT)^2}\right]}\,,\label{cha}\end{eqnarray}
showing that it ultimately decreases uniformly as $(J/kT)^{2}$ in this limit.
It then exhibits a reentry for $T>0$ in all $T=0$ separable sectors, as seen in
Figs.\ \ref{f5}  and \ref{f6}.

As previously mentioned, by applying sufficiently strong opposite fields at
each site it is possible to effectively ``cool down'' the thermal state
$\rho_{12}$ at any $T>0$,  bringing it as close as desired to the entangled
state $|\Psi^-\rangle\langle\Psi^-|$. This behaviour is shown in the right
panel of Fig.\ \ref{f5}. It is seen  that the entanglement of formation $S$, 
obtained from the thermal concurrence $C$ by the same expression (\ref{S})
\cite{WW}, and the relative entropy of coherence, initially different and
dependent on $J_z$, merge for increasing values of $|h_1-h_2|$,  approaching a
common $J_z$-independent limit which  is the entanglement entropy $S$  of the
pure state $|\Psi^-\rangle$.  The vanishing difference between $S$ and
$C_h$ for high $|h_1-h_2|$ is a clear signature that  $\rho_{12}$ has become
essentially pure.

\section{The spin-$1$ pair}
We now consider the $s=1$ case.  The behaviour is essentially similar to that
for   $s=1/2$, the main difference being the appearing of an {\it intermediate
$M=\pm 1$  magnetization step} in the $T=0$ diagrams, between the entangled
$M=0$ GS and the aligned separable $M=\pm 2$ states. This effect leads to an
entanglement step since the $M=\pm 1$ GS is less entangled than the $M=0$ GS.

Using now the notation $|m_1,m_2\rangle$ for the states of the standard basis,
with $m_i$ the eigenvalues of $s_i^z$, the GS of the spin $1$ $XXZ$ pair can be
one of the $|M|=2$ aligned states $|\Psi_{\pm 2}\rangle=|\pm 1,\pm 1\rangle$,
one of the $|M|=1$ states, which will be of the form

\begin{equation}
|\Psi_{\pm 1}\rangle=\cos\alpha_{\pm}|\pm 1,0\rangle+\sin\alpha_{\pm}|0,\pm 1\rangle\,,
\end{equation}
with $\tan\alpha_{\pm}=\pm\frac{\eta}{2}-\sqrt{1+\frac{\eta^2}{4}}$ and
$\eta=\frac{h_1-h_2}{J}$, and one of the $M=0$ states, of the form
\begin{equation}
|\Psi_{0}\rangle=\gamma_+|1,-1\rangle+\gamma_0|00\rangle+\gamma_{-}|-1,1\rangle\,.
\end{equation}`
All  coefficients are independent of $h_1+h_2$, but those of $|\Psi_0\rangle$
depend now on $J_z$. For $J_z=0$  they can be still written down concisely:
$\gamma_0/\gamma_+=\eta-\sqrt{2+\eta^2}$,
$\gamma_-/\gamma_+=1+\eta\gamma_0/\gamma_+$.

\begin{figure}[htb]
    \centerline{\includegraphics[scale=0.7]{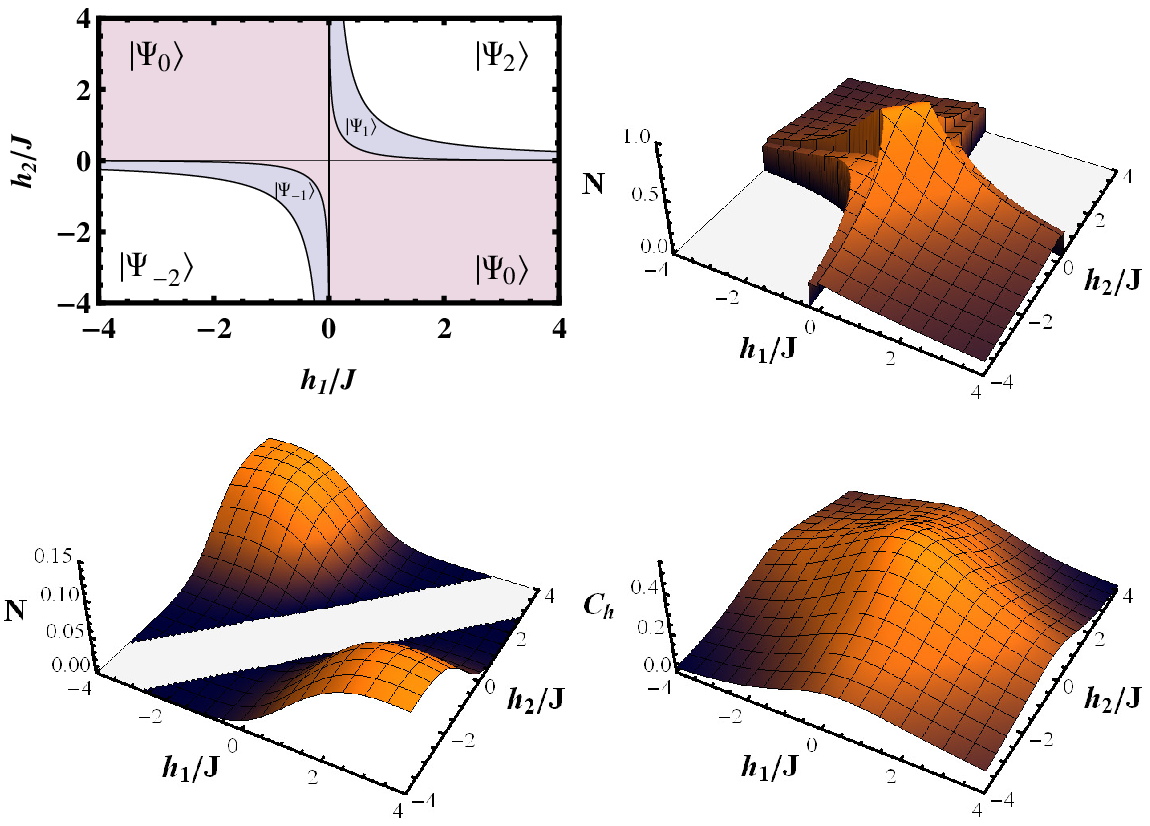}}
\caption{Top: Ground state phase diagram (left) and negativity (right) for the
spin $1$ pair with $J_z=0$ as a function of the applied fields at each spin.
Colored sectors on the left indicate entanglement. Bottom: The negativity
(left) and relative entropy of coherence (right) of the thermal state of the
spin $1$ pair at temperature $kT=J$ as a function of the applied fields.}
\label{f7}
\end{figure}

Their energies are
\begin{eqnarray} E_{\pm 2}&=&\mp(h_1+h_2)+J_z\,,\nonumber\\
E_{\pm 1}&=&{\textstyle\mp\frac{h_1+h_2}{2}-\sqrt{J^2+(\frac{h_1-h_2}{2})^2}}\,,\\
E_0&=&-\sqrt{2J^2+(h_1-h_2)^2}\;\;\;\;\;\;\;(J_z=0)\,.
\end{eqnarray}
The border of the $T=0$ entangled region, determined by that between the
$|M|=2$ and $|M|=1$ GS,  $E_{\pm 2}=E_{\pm 1}$, is then given again by Eqs.\
(\ref{c0})--(\ref{c2}) with $J\rightarrow 2J$, $J_z\rightarrow 2J_z$. The GS
phase diagrams have then the same forms as those of Fig.\ \ref{f1} except for
the previous rescaling and the magnetization step. The $J_z=0$ case is shown in
Fig.\ \ref{f7}.

As entanglement measure valid for both zero and finite temperature, we will now use 
the negativity \cite{VW.02,ZHSL.98}, a well-known entanglement monotone which is 
computable in any mixed state, since an explicit expression for the concurrence or 
entanglement of formation of a general mixed state of two qutrits (spin $1$ pair) or 
in general two qudits with $d\geq 3$ is no longer available.  The negativity 
is minus the sum of the negative eigenvalues of the partial transpose
\cite{P.96,HHH.96} $\rho_{12}^{\rm t_2}$ of $\rho_{12}$:
\begin{equation}
N(\rho_{12})=({\rm Tr}|\rho_{12}^{\rm t_2}| -1)/2\,.
\end{equation}
A non-zero negativity  implies entanglement, whereas for mixed states, the
converse is not necessarily true (except for  two qubit states \cite{P.96,HHH.96} 
or special states),  vanishing for bound entangled states. Nonetheless it is normally used
as an indicator of useful entanglement.

For pure states it reduces to a special entanglement entropy \cite{RC.05}, 
being a function of  the one-spin reduced state $\rho_1$ (or $\rho_2$, isospectral with $\rho_1$
for a pure state):    $N=\frac{1}{2}[({\rm Tr}\,\sqrt{\rho_1})^2-1]$. It is
then non-zero if and only if  the state is entangled. Its  maximum value for a spin $s$
pair is $N=s$. We then obtain $N(|\Psi_{\pm 2}\rangle)=0$,
\begin{eqnarray}
N(|\Psi_{\pm 1}\rangle)&=&{\textstyle\frac{1}{2}|\sin(2\alpha_{\pm})|=\frac{1}{\sqrt{4+\eta^2}}}\,,\\
N(|\Psi_{0}\rangle)&=&{\textstyle\frac{1}{2}}[|\gamma_+\gamma_-|+|\gamma_0|
(|\gamma_+|+|\gamma_-|)]\\
&=&{\textstyle\frac{1+\sqrt{2}
 \sum\limits_{\nu=\pm}\sqrt{1+\eta(\eta+\nu\sqrt{2+\eta^2})}}{2(2+\eta^2)}}
\;\;\;\;\;(J_z=0)\,.
\end{eqnarray}
They are decreasing functions of $\eta=\frac{h_1-h_2}{J}$, reaching for
$\eta=0$ the value $\frac{1}{2}$ for $|M|=1$ (maximum value for Schmidt rank
$2$)  and $\frac{1+2\sqrt{2}}{4}\approx 0.958$ for  $M=0$. The negativity of
the $M=0$ GS depends now on $J_z$, reaching the maximum $N=1$  for $J_z=1$.

It is verified in Fig.\ \ref{f7} that the $T=0$ phase diagram and entanglement
for $J_z=0$ is similar to that for $s=1/2$ except for the $M=\pm 1$
magnetization and negativity steps. Remarkably, the finite temperature
negativity  diagram {\it is again characterized by a separability stripe}
$|h_1-h_2|\leq h_c(T,J,J_z)$ for $T>T_c$, with  the boundary of the non-zero
negativity  sector  {\it independent of $h_1+h_2$} (as demonstrated in the next
section). At $J_z=0$ the stripe emerges for  $kT>kT_c\approx 0.864 J$ (root of
the critical equation $3+2\cosh(2\beta_c J)=\cosh(2\sqrt{2}\beta_c J)$), with
the whole field plane entangled for $T<T_c$. It should be also mentioned that
the $T=0$ negativity step gives rise to a negativity ``valley'' for low finite
$T$  due to the convexity of $N$, as will be seen in the  next section.

The relative entropy of coherence in the standard basis behaves in the same way
as before. It approaches the GS entanglement entropy for $T\rightarrow 0$,
while for $kT\gg J, J_z, |h_1|, |h_2|$, it decreases uniformly at leading
order, becoming, for $J_z=0$, 
\begin{equation} C_h(\rho_{12})\approx {\textstyle\frac{4}{9\ln 2}(\frac{J}{kT})^2[1-
    \frac{55 J^2+15(h_1+h_2)^2+9(h_1-h_2)^2}{(12 kT)^2}]}\,.\label{cap1}
\end{equation}

\section{The spin $s$ case}
\subsection{Ground state phase diagram and entanglement}
\begin{figure}[htb]
    \centerline{\includegraphics[scale=1.1]{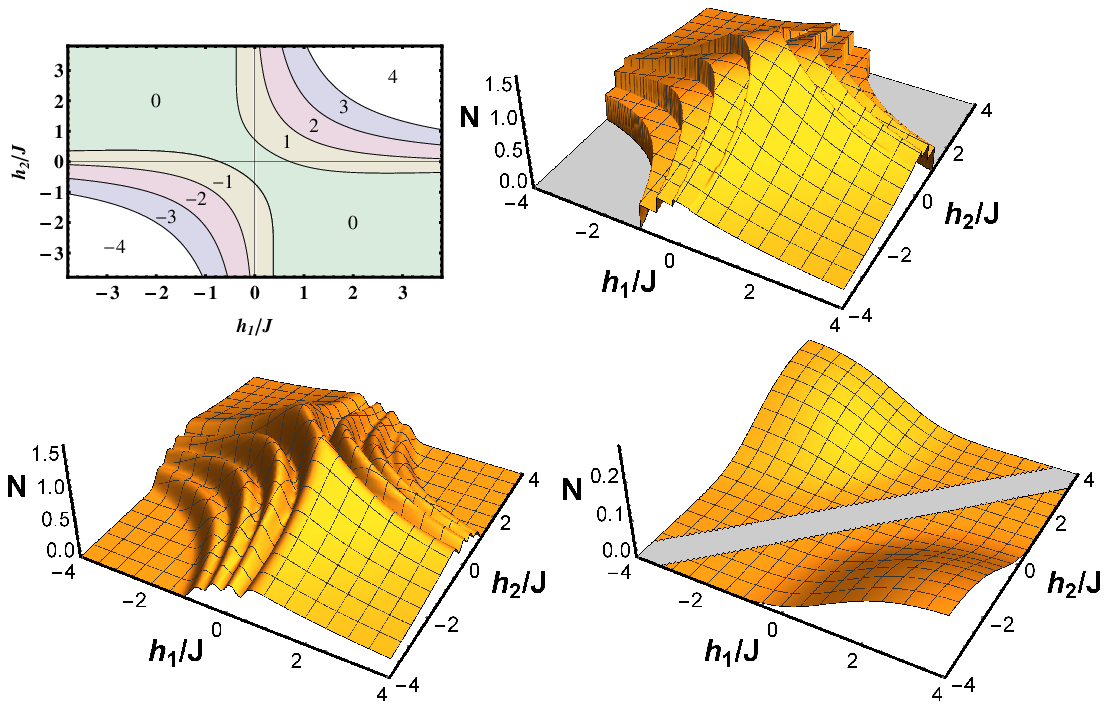}}

\caption{Top: Ground state phase diagram (left) and negativity (right) for a
spin $2$ pair with $J_z=0$ as a function of the applied fields at each spin.
Colored sectors on the left indicate entanglement, with the number denoting the
magnetization of the state. Bottom: The negativity of the spin 2 pair at
temperature $kT=0.1 J$ (left) and $kT=1.6 J$ (right).} \label{f8}
 \end{figure}

Let us finally consider the main features of the general spin $s$ $XXZ$ pair in
a non-uniform field. The GS phase diagram remains similar to the previous
cases, but now with {\it  $2s$ magnetization steps}, from  $M=0$ up to $M=\pm
2s$. These steps originate $2s$ steps in the $T=0$ entanglement  and
negativity, since they  decrease with increasing $|M|$. This behaviour can be
seen in the top panels of Fig.\ \ref{f8} for an $s=2$ pair. \\

\vspace*{-0.5cm}

The border of the entangled region in the field plane is determined  by that
between the aligned  GS with $|M|=2s$ and the entangled GS with $|M|=2s-1$.
Remarkably, {\it it  is  the same as that for $s=1/2$, Eq.\ (\ref{c0})}, with
the rescaling $J\rightarrow 2sJ$, $J_z\rightarrow 2s J_z$:
\begin{equation}
|h_1+h_2|<2sJ_z+\sqrt{4s^2J^2+(h_1-h_2)^2}\,.\label{crits}
\end{equation}
The border are then the hyperbolas (\ref{c1})--(\ref{c2}) with the previous
scaling and give rise to the same possibilities depicted in Fig.\ \ref{f1},
with the additional inner magnetization steps.\\
{\it Proof:} Considering first $h_1+h_2\geq 0$, the energies of the $M=2s$
aligned state $|ss\rangle$ and the lowest  $M=2s-1$ state, which is
$|\Psi_{2s-1}\rangle=\cos\alpha|s,s-1\rangle+\sin\alpha|s-1,s\rangle$, with
$\tan\alpha=\frac{\eta}{2s}-\sqrt{1+\frac{\eta^2}{4s^2}}$ and
$\eta=\frac{h_1-h_2}{J}$,  are
\begin{eqnarray}
E_{2s}&=&-{\textstyle s(h_1+h_2)+s^2J_z}\,,\label{E2s}\\
    E_{2s-1}&=&{\textstyle-(2s-1)\frac{h_1+h_2}{2}+s(s-1)J_z-
    	\sqrt{(\frac{h_1-h_2}{2})^2+s^2 J^2}}
    \,.\label{E2sm1}\end{eqnarray}
The condition $E_{2s-1}<E_{2s}$ leads then to Eq.\ (\ref{crits}). If
$h_1+h_2\leq 0$, the result is similar with $h_1+h_2$ replaced by $|h_1+h_2|$
and $E_{M}$ by $E_{-M}$.  \qed

 \begin{figure}[htb]
    \centerline{\includegraphics[scale=1.1]{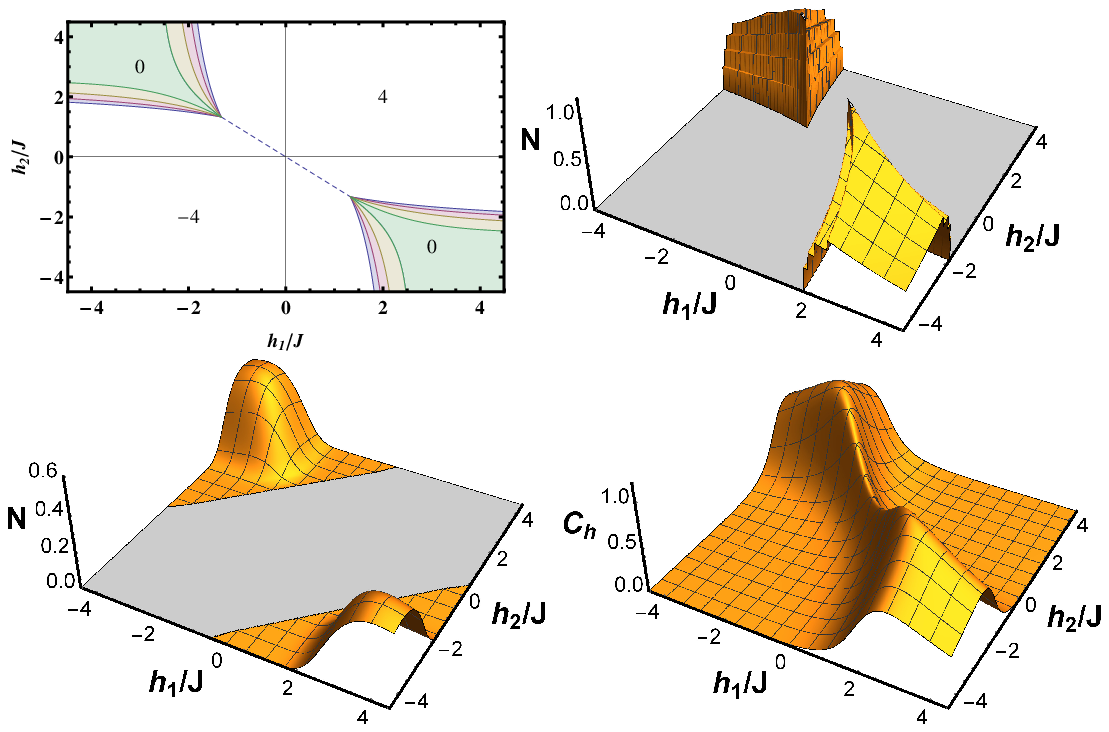}}
\caption{Top: Ground state phase diagram (left) and negativity (right) for a
spin $2$ pair with $J_z=-1.2J$ as a function of the applied fields at each
spin. Colored sectors indicate an entangled GS, with the color and number
identifying the distinct magnetizations. Bottom: The negativity (left) and
relative entropy of coherence (right) of the spin 2 pair for $J_z=-1.2J$ at
temperature $kT=0.5J$. } \label{f9}
 \end{figure}

In Fig.\ \ref{f9} we plot an example for $s=2$ of the interesting FM case
$J_z<-J$, where the GS is fully aligned for any uniform field $h_1=h_2=h$, as
in the $s=1/2$ case,  with a transition $M=-2s$ to $M=2s$ at $h=0$. However,
{\it it  can again be entangled with a non-uniform field}, by applying {\it
opposite} fields at each spin. Eq.\ (\ref{crits}) implies that in this case GS
entanglement will arise for
 \begin{equation} |h_1-h_2|>h_c=2s\sqrt{J_z^2-J^2}\,\;\;\;\;\;(J_z<-J)\,,\end{equation}
within the limits (hyperbolas) determined by Eq.\ (\ref{crits}), which entail
that no entangled GS will arise  for fields of equal sign if $J_z<-J$, as in
the $s=1/2$ case. Moreover, the edges of the $T=0$ entangled sector,
 \begin{equation}h_1=-h_2=\pm h_c/2\,,\end{equation}
are actually   {\it critical points}  in which $4s+1$ distinct GS's,
corresponding to all magnetizations  $M=-2s,\ldots, 2s$, {\it coalesce and
become degenerate}, as verified in the top left panel of Fig.\ \ref{f9}. At
these points, their  common energy is
 \begin{equation}
 E_M=s^2 J_z<0\;\;\;\;(h_1=-h_2=\pm h_c/2,\;J_z<-J,\;M=-2s,\ldots, 2s),
 \end{equation}
independent  of $J$ and $M$. Their entanglement  decreases, however, with $|M|$,
as seen in the top right panel through the negativity. Along the line
$h_1=-h_2$ a GS transition from the lowest non-degenerate $M=0$ state to the
aligned states $M=\pm 2s$  (degenerate along this line) occurs at $h_1=-h_2=\pm
h_c/2$,  although precisely at these points the GS becomes $4s+1$-fold
degenerate. Actually the transition region with intermediate GS magnetizations
$|M|=1,\ldots,2s-1$ is rather narrow in the $h_1,h_2$ field plane, as seen in
the top left panel, collapsing at the critical points.

A final comment is that the maximum GS entanglement  of a spin $s$ $XXZ$ pair
is reached at the $M=0$ GS and depends on $J_z$ for $s>1/2$. For $J_z<-J$ it is
reached at the previous critical points (Fig.\ \ref{f9}), while for $J_z>-J$ it
is reached  along the line $h_1=h_2$ (Fig.\ \ref{f8}).  In the uniform AFM case
$J_z=-J$, the $M=0$ eigenstate will be {\it maximally entangled} for $h_1=h_2$,
leading to maximum negativity $N=s$, while for $-J<J_z<J$, the $M=0$ negativity
will be smaller and proportional to $\sqrt{s}$ for large $s$, due to a gaussian
profile of width $\propto \sqrt{s}$ of the expansion coefficients in the
standard basis \cite{BRCM.16}. We also mention that some internal magnetization
steps may disappear for large $J_z>J$ and small $|h_1-h_2|$.

   \subsection{Finite temperatures}
As $T$ increases, the $T=0$ negativity steps become  initially negativity {\it
valleys}, as clearly seen in the bottom left panel of Fig.\ \ref{f8}, since
convexity of $N$ implies that its value for the mixture of two entangled states
will be smaller than the average negativity of the states.  These valleys are
rapidly smoothed out as $T$ increases further.  On the other hand,  it is also
seen  in Figs.\ \ref{f8}--\ref{f9}  that entanglement  diffuses outside the
$T=0$ entangled region as $T$ increases, covering initially the whole field
plane for $J_z>-J$ and the whole plane outside the stripe $|h_1-h_2|\leq h_c$
for $J_z<-J$, although the negativity will be small in the $T=0$ aligned
sectors.

A striking feature for finite  temperatures is the {\it persistence of a
separability stripe} $|h_1-h_2|\leq h_c(T,J,J_z)$ in the field plane when
considering the thermal entanglement,  as seen in the bottom right panel  of
Fig.\ \ref{f8} for $J_z=0$,  where the stripe emerges for $T>T_c=1.498 J/k$,
and in the bottom left panel of Fig.\ \ref{f9} for $J_z=-1.2 J$, where the
stripe is present $\forall$ $T>0$. This result will now  be shown to hold for
arbitrary spin, following the arguments of \cite{RC.05} for  $XXZ$
systems in uniform fields.

{\it Lemma 1. The limit condition  for entanglement and 
	non-zero negativity of a spin $s$ pair
with an $XXZ$ coupling in a non-uniform transverse field at temperature $T>0$,
depends only on the  field difference $h_1-h_2$. This result applies also to
any coupling independent of the field that commutes with the total spin along
$z$ ($[H,S_z]=0$).}

 {\it Proof:}
 We first rewrite the Hamiltonian of a spin $s$ pair in a non-uniform field as
 \begin{eqnarray}
 H&=&-\frac{h_1+h_2}{2}(s^z_{1}+s^z_{2})-\frac{h_1-h_2}{2}(s^z_{1}-s^z_{2})+V\label{H2}
 \end{eqnarray}
 where $V$ denotes  the (field independent) interaction between the spins, assumed to satisfy
 $[V,S_z]=0$ ($S_z=s^z_1+s^z_2$). The first term in (\ref{H2}) is the uniform field component
 and commutes with the rest of the Hamiltonian. Consequently, the thermal state for average field
 $h=\frac{h_1+h_2}{2}$, $\rho_{12}(h)=Z_h^{-1}e^{-\beta H}$,  can be written as
 \begin{equation}\rho_{12}(h)=
 {\textstyle\frac{Z_0}{Z_h}}e^{\beta h S_z/2} \rho_{12}(0)e^{\beta h S_z/2}\,,\label{r0}
 \end{equation}
where $\rho_{12}(0)$ depends just on $h_1-h_2$ and commutes with $S_z$. 

Eq.\ (\ref{r0}) implies that $\rho_{12}(h)$ will be separable, i.e., a convex
combination of   product states \cite{W.89}, {\it  if and only
	if $\rho_{12}(0)$ is separable}: If $\rho_{12}(0)=\sum_{\alpha}
q_\alpha \rho_1^\alpha\otimes \rho_2^\alpha$, with  $q_\alpha>0$ and $\rho_i^\alpha$ 
local mixed states, then
$\rho_{12}(h)=\sum_{\alpha}q_\alpha
\tilde{\rho}_1^\alpha\otimes\tilde{\rho}_2^\alpha$ with
$\tilde{\rho}_i^\alpha\propto e^{\beta h s^z_i/2}\rho_i^\alpha e^{\beta h
	s^z_i/2}$ also local mixed states, so that it is separable as well. Similarly, 
$\rho_{12}(h)$ separable implies $\rho_{12}(0)$  separable 
($\rho_i^\alpha\propto e^{-\beta hs^z_i/2}\tilde{\rho}_i^\alpha e^{-\beta h s^z_i/2}$). 
Hence, the limit condition for exact separability  depends only on $h_1-h_2$. 

Let us now consider the negativity. The non-zero matrix elements of $\rho_{12}(h)$ are $\langle m,M-m|\rho_{12}(h)|M-m',m'\rangle\propto
e^{\beta hM} \langle m,M-m|\rho_{12}(0)|M-m',m'\rangle$. Its partial transpose
will then have  matrix elements $\langle m,m'|\rho_{12}^{\rm
t_2}(h)|M-m',M-m\rangle\propto e^{\beta h M}\langle m,m'|\rho_{12}^{\rm
t_2}(0)|M-m',M-m\rangle$, such that  it can also be written as
 \begin{equation}
 \rho_{12}^{t_2}({h})={\textstyle\frac{Z_0}{Z_h}}
 e^{\beta {h}S_{z}/2}\rho_{12}^{t_2}(0)e^{\beta {h}S_{z}/2}\,.
 \label{t2}
 \end{equation}
Although $\rho_{12}^{t_2}(0)$ will no longer commute with $S_z$,
$\rho_{12}^{t_2}(h)$ will be positive definite (i.e., with positive
eigenvalues) if and only if $\rho_{12}^{t_2}(0)$ is  positive definite,   since
$e^{\beta hS_z/2}$ is positive definite and $Z_h$, $Z_0$ are positive. This
result  demonstrates the Lemma for the negativity. More explicitly, the onset for non-zero
negativity occurs when the lowest eigenvalue of $\rho_{12}^{t_2}$ becomes
negative, implying a vanishing eigenvalue at the onset, i.e., ${\rm det}
[\rho_{12}^{t_2}]=0$. But Eq.\ (\ref{t2}) implies ${\rm det}
[\rho_{12}^{t_2}(h)]= (\frac{Z_0}{Z_h})^{(2s+1)^2}{\rm det}
[\rho_{12}^{t_2}(0)]$ (as ${\rm Tr}\, S_z=0$), so that the  critical conditions
at $h\neq 0$ and $h=0$ are equivalent.
 \qed

In addition, for an $XXZ$ coupling  as well as for any coupling invariant under
permutation of the spins, the limit condition will obviously depend only on the
{\it absolute value}  $|h_1-h_2|$ of the field difference, as those for
$(h_1,h_2)$ and $(h_2,h_1)$ should be identical.

Therefore, even though the negativity for the $XXZ$ pair does depend on the
average field $h=\frac{h_1+h_2}{2}$ (through the relative weights of the
distinct eigenstates), as was seen in previous figures, the limit temperature
for non-zero negativity at fixed exchange couplings, and the threshold values
of $J_z$ or $J$ for non-zero negativity at fixed temperature,  will depend just
on $|h_1-h_2|$. In the $h_1,h_2$ field plane, the set of zero negativity states
will then be stripes, i.e., typically a stripe $|h_1-h_2|\leq h_c$. Of course,
$N$ can be exponentially small outside the stripe, but not strictly zero.

The previous features of  the relative entropy of coherence remain also valid.
The standard basis of states $|m_1,m_2\rangle$ continues to be the Schmidt
basis for definite magnetization eigenstates, i.e. $|\Psi_M\rangle=\sum_{m}c_m
|m,M-m\rangle$, entailing that for $T\rightarrow 0$ the coherence will approach
the GS entanglement entropy (for a non-degenerate GS) adopting qualitatively 
the same form as the $T=0$ negativity.
Nevertheless, as $T$ increases the $T=0$  steps will become rapidly smoothed
out in the coherence, without exhibiting minima or valleys. It will also
rapidly occupy the $T=0$ separable sectors, becoming in particular  prominent
along the line $h_1=-h_2$ for $J_z<-J$, as seen in the bottom right panel of
Fig.\ \ref{f9}. On the other hand,  for sufficiently high temperatures it will
approach  a uniform decay pattern for all $s$. A series expansion for $kT\gg
J,|J_z|, |h_1|, |h_2|$ leads to
\begin{equation}
C_h(\rho_{12})\approx \frac{\beta^2{\rm Tr}[H^2-(H_{\rm
diag})^2]}{2 d\ln 2}=\frac{1}{9\ln 2} \left(\frac{s(s+1)J}{kT}\right)^2\,,\label{cas}
\end{equation}
where the first result holds in a system of finite dimension $d$ and the last
one is the leading asymptotic expression for a spin $s$ $XXZ$ pair.  It
reproduces the leading term of previous asymptotic results (\ref{cha}) and
(\ref{cap1}).

\section{Conclusions}
We have discussed in  detail the entanglement and coherence of the $XXZ$  spin
$s$ pair in a transverse non-uniform field  at both zero and finite
temperatures. The general spin $s$ case exhibits interesting features which can
already be seen in the basic $s=1/2$ case. In the latter,  while the $T=0$
diagram in the field plane is characterized by an $M=0$ entangled region
bounded by hyperbola branches,  reachable through non-uniform fields even in
the FM case $J_z<-J$,  the thermal state is  characterized by a {\it
separability stripe} $|h_1-h_2|\leq h_c(T,J,J_z)$ in the $h_1, h_2$ field plane
for {\it any} $T>0$, with the system becoming pure and entangled for large
values of $|h_1-h_2|$.  Analytic expressions were provided.

Remarkably, these features were shown to remain strictly valid for any value of
the spin $s$. The boundaries of the $T=0$ entangled sector are given by the
same expressions with a simple rescaling, while the conditions for non-zero
thermal entanglement and negativity were rigorously shown to depend just on the
field difference $|h_1-h_2|$ for any $s$, entailing that for $T>0$ strict
separability will still be restricted to a stripe $|h_1-h_2|\leq h_c(T,J,J_z)$. The
main difference with the spin $1/2$ case is the emergence of $2s$ magnetization
and entanglement steps  at $T=0$, which lead to deep valleys in the negativity
at low temperatures but which disappear as $T$ increases. Another interesting
aspect emerging for non-uniform fields for increasing spin is the appearing of
a critical point along the line $h_1=-h_2$, which determines the onset of GS
entanglement for  $J_z<-J$ and where all $4s+1$ GS's with magnetizations
$M=-2s,\ldots,2s$ coalesce.

The relative entropy of coherence in the standard basis approaches 
the entanglement entropy for $T\rightarrow 0$, although for $T > 0$ it stays
non-zero for all fields. The exact asymptotic expression for high $T$ was
derived, which shows that it ultimately decays uniformly as $(s(s+1)J/kT)^{2}$
for sufficiently high temperatures.

In summary, the present results show that the $XXZ$ pair in a non-uniform field
is an attractive  simple system with potential for quantum information applications. 
Its entangled eigenstates, having definite magnetization, admit a variable degree of
entanglement  which can be controlled by tuning the fields at each spin.
Moreover such tuning enables to  select the magnetization of the GS at $T=0$
for any anisotropy, while at  $T>0$ it allows one to effectively cool down the
system to an entangled state. At $T=0$ entanglement itself can be detected and
approximately measured through the magnetization, since it decreases with
increasing $|M|$ and vanishes just for maximum $|M|$.  The possibility of
simulating $XXZ$ systems with tunable couplings and fields by different means
enhances the interest in this type of models. It would then be interesting to 
extend these results to spin $s$ $XXZ$ chains and explore in detail their entanglement 
and coherence properties under non-uniform fields. Preliminary results indicate that 
at least for small $n$,  the general behavior of an $n$-spin-$s$ chain in a  general  
field does resemble that of an effective spin pair with the same total maximum spin 
(i.e., a spin $ns/2$ pair), although details depend on several features like boundary 
conditions, parity of $n$, etc. (and in the case of entanglement and coherence, of course 
on the type of pair or partition analyzed), which are currently under investigation.

\section{Acknowledgments}
This work was supported in part by the Departamento de Ingenier\'ia Qu\'imica
UTN-FRA (ER), Comisi\'on de Investigaciones Cient\'{\i}ficas (CIC) (RR), 
and CONICET (NC) of Argentina. Authors also acknowledge support from  CONICET 
grant PIP 11220150100732.

\end{document}